\documentclass[twocolumn]{revtex4-1}
\usepackage{amsmath,amssymb,amsfonts}
\usepackage{mathtools}
\usepackage{multirow}
\usepackage{graphicx}
\usepackage{dcolumn}
\usepackage{bm}
\usepackage{footmisc}
\usepackage[english]{babel}
\usepackage{natbib,hyperref}
\usepackage[normalem]{ulem}
\usepackage{comment}
\usepackage{xcolor}
\usepackage{caption} 
\usepackage{booktabs} 

\preprint{APS/123-QED}
\begin{document}
\renewcommand{\arraystretch}{1.75}

\title{Rapid Statistical-Physical Adversarial Downscaling\\Reveals Bangladesh's Rising  Rainfall Risk in a Warming Climate }

\author{\begin{tabular}{ccc}
Anamitra Saha && Sai Ravela\\
anamitra@mit.edu && ravela@mit.edu\\
&Earth Signals and Systems Group&\\
\end{tabular}\\
Earth, Atmospheric and Planetary Sciences \\ 
Massachusetts Institute of Technology\\
Cambridge, MA, USA 
}

\thispagestyle{plain}
\pagestyle{plain}

\sloppy
\begingroup
\hyphenpenalty 10000

\begin{abstract}
In Bangladesh, a nation vulnerable to climate change, accurately quantifying the risk of extreme weather events is crucial for planning effective adaptation and mitigation strategies. Downscaling coarse climate model projections to finer resolutions is key in improving risk and uncertainty assessments. This work develops a new approach to rainfall downscaling by integrating statistics, physics, and machine learning and applies it to assess Bangladesh's extreme rainfall risk. Our method successfully captures the observed spatial pattern and risks associated with extreme rainfall in the present climate. It also produces uncertainty estimates by rapidly downscaling multiple models in a future climate scenario(s). Our analysis reveals that the risk of extreme rainfall is projected to increase throughout Bangladesh mid-century, with the highest risk in the northeast. The daily maximum rainfall at a 100-year return period is expected to rise by approximately 50 mm per day. However, using multiple climate models also indicates considerable uncertainty in the projected risk.
\end{abstract}

\maketitle

\section{Introduction}
The rising trend of extreme weather events is a pressing concern for Bangladesh, a developing nation heavily exposed to climate-related risks~\cite{eckstein2021global}. Increased frequencies of tropical cyclones, extreme rainfall, floods, and heatwaves severely threaten food security, water resources, public health, and the country's critical infrastructure. Given the substantial losses to lives and livelihoods that Bangladesh has experienced due to these events in the past, it is crucial to evaluate how the risks associated with them may evolve as the climate continues to change in the coming decades.

Accurate risk assessment of weather extremes requires high-resolution local information. While such data can be acquired from local gauge measurements and satellite observations in the present climate, future risk assessment must rely on climate projections, especially using numerical climate models at decadal intervals and beyond. However, climate models are computationally expensive, with operational throughput for state-of-the-art kilometer-scale global models falling short of the targeted 1 simulated year per wall-clock day (SYPD)~\cite{schulthess2018reflecting}. Even if this benchmark were achieved, producing an ensemble of simulations across multiple models, parameters, and climate scenarios would still require several years. Handicapped by their inefficiency and limited regional focus, climate models usually produce output at a much coarser resolution. Consequently, finer-scale geophysical processes are not sufficiently represented, leading to bias and underestimation of risks. Thus, downscaling coarse model outputs to finer resolutions is particularly interesting.

One approach, namely dynamical downscaling, nests numerical models of theory to simulate climate at a finer resolution within a limited (regional) domain of a much coarser global climate model~\cite{prein2020simulating}. While this approach is more resource-efficient than running a high-resolution model for the entire globe, the computational cost of running multiple such simulations is still prohibitively high, limiting their practicality in robustly quantifying risk and, more importantly, its uncertainty. Additionally, models often do not match the observations, whether run at fine resolution globally or downscaled dynamically regionally. There are timing errors and pattern mismatches, so bias corrections become central, pointwise comparisons are complicated, and risk assessments must become distributions-oriented to incorporate these uncertainties. 

 In contrast, data-driven techniques, including statistical or machine-learning (ML) models, offer a more efficient approach to downscaling and risk quantification. These approaches view downscaling as an ill-posed inverse problem in which a single coarse-resolution input can produce a large ensemble of plausible high-resolution solutions. 
 
 However, such methods also have some inherent limitations. While prolific at generating solutions, not all downscaling solutions will match the high-resolution observed reality.  ML-downscaled solutions may be physically unrealistic because they typically lack an intrinsic understanding of the underlying physics and may not even adhere to basic physical principles without additional support. Further, extreme events are rare by nature, making them challenging to model with supervised ML techniques that are invariably data-hungry. To complicate matters further, in the climate model downscaling scenario, low-resolution data typically comes from climate models, and high-resolution training data can come from observations or disparate models. In practice, this means a lack of one-to-one correspondence, as previously noted in timing and pattern errors, which presents a significant complication. Thus, learning must often incorporate distributions-oriented measures from theoretical frameworks or empirical data in addition to or instead of pixelwise measures.

Saha and Ravela (2024), hereafter referred to as SR24, propose an approach that combines statistics, simplified physics, and adversarial learning to efficaciously downscale extreme precipitation and overcome some of the limitations mentioned above~\cite{saha2024statistical}. SR24 is suitable for rapidly quantifying the risk of extremes and their uncertainties. Still, as a proof of concept, it has only been applied to downscale urban-scale climate model precipitation in the present climate. This study introduces several enhancements to the method and extends the application to future climate scenarios at the national scale. In particular, the downscaling model is refined to capture better spatial heterogeneity (texture) of high-resolution rainfall data. We also improve risk and uncertainty assessment with a progressive mixture of distributions for extreme and non-extreme ranges. Our model is applied to assess Bangladesh's current and future risk of extreme precipitation. Results mid-century indicate that the risk is expected to increase nationwide, especially in the Northeast. However, there remains considerable uncertainty in model projections, which we posit is essential to communicate for informing adaptation and mitigation policy decisions in Bangladesh.

The remainder of the manuscript is organized as follows: Section~\ref{section:relatedworks} reviews prior research on downscaling. Section~\ref{section:methods} describes the data and the methodology. Section~\ref{section:results} presents our findings. Finally, section~\ref{section:discussions} discusses this study's significance, limitations, and future scope.

\section{Related Work}
\label{section:relatedworks}
Downscaling approaches can be broadly categorized into two paradigms: theory-driven and data-driven. A hybrid approach combining both techniques is also common. The primary theory-driven method is dynamical downscaling, which uses regional climate models to simulate higher-resolution data within a limited area. While several regional models, such as Climate Limited-area Model (CLM)~\cite{sorland2021cosmo}, Weather Research and Forecasting model (WRF)~\cite{katragkou2015regional}, and Model for Prediction Across Scales (MPAS)~\cite{sakaguchi2023technical} have been widely employed for downscaling and climate change impact assessments, their biggest drawback is computational overhead. Alternative simpler physics-based models, like the upslope model~\cite{roe2005orographic} and the linear model for orographic precipitation (spectral method)~\cite{smith2004linear}, offer less computational demand. Such models have also been used for precipitation downscaling but require significant statistical post-processing due to incomplete physics~\cite{paeth2017efficient}.

On the other hand, data-driven statistical approaches have gained traction as observational data become more abundant. Techniques such as bias correction and spatial disaggregation~\cite{wood2002long}, multi-linear regression~\cite{sachindra2014multi}, generalized linear model~\cite{beecham2014statistical}, kernel density estimators~\cite{lall1996nonparametric}, kernel regression~\cite{kannanNonparametricKernelRegression2013}, Markov model~\cite{mehrotra2005nonparametric} and Bayesian model averaging~\cite{zhang2015new} have been used for rainfall downscaling. Machine learning methods such as Random Forest~\cite{heSpatialDownscalingPrecipitation2016}, Support Vector Machine~\cite{tripathiDownscalingPrecipitationClimate2006}, Relevant Vector Machine~\cite{george2023multi}, and genetic programming~\cite{sachindraMachineLearningDownscaling2019} have also been applied.

Various deep learning models, such as Recurrent Neural Networks (RNNs)~\cite{wangSequencebasedStatisticalDownscaling2020}, Long Short-Term Memory (LSTM) networks~\cite{tran2019downscaling}, and autoencoders~\cite{vandalIntercomparisonMachineLearning2019} have been leveraged for rainfall downscaling. These methods excel at extracting features from complex climate datasets due to their deep structures. Convolutional Neural Network (CNN)-based techniques, initially developed for image super-resolution, have also been adapted for climate science to capture spatial structures in climate variables~\cite{reddy2023precipitation, vandal2017deepsd}. For example, Super-resolution CNN (SRCNN) was an early successful model for downscaling~\cite{dong2015image}, followed by others like Very Deep Super-resolution (VDSR)~\cite{kim2016accurate}, Enhanced Deep Super-resolution (EDSR)~\cite{lim2017enhanced}, Laplacian Pyramid Super-Resolution Network (LapSRN)~\cite{lai2017deep} and Multi-Scale Residual Block (MSRB)~\cite{li2018multi}. Generative Adversarial Network (GAN)-based models, such as Super-resolution GAN (SRGAN)~\cite{ledig2017photo} and Enhanced Super-resolution GAN (ESRGAN)~\cite{wang2018esrgan} improved upon them~\cite{saha2024statistical,hess2022physically,wang2021fast}. UNET~\cite{sha2020deep,sharma2022resdeepd}, Normalizing Flows~\cite{groenke2020climalign,winkler2024climate}, Attention mechanism~\cite{xiang2022novel}, Diffusion models~\cite{ling2024diffusion} have also been applied to the downscaling problem.

Several downscaling techniques have been employed for downscaling precipitation in Bangladesh~\cite{islam2022future,wu2022statistical,pour2018model}. However, these studies primarily focus on mean and seasonal rainfall climatology of the rainfall, and the resolution of these results remains relatively coarse. Recognizing the need for very high-resolution downscaled rainfall data across the country for extreme event risks in present and future climates, we have adapted SR24~\cite{saha2024statistical} in this study.

\section{Data and Methods}
\label{section:methods}
\subsection{Study Region}
Bangladesh, a low-lying coastal county in Southeast Asia, is vulnerable to climate change due to its high population density. Bangladesh's diverse topography and climate zones create complex spatial and temporal rainfall patterns. The terrain of Bangladesh largely consists of the alluvial plains of the Ganges, Brahmaputra, and Meghna rivers, forming the world's largest delta. This deltaic region experiences significant hydrological variability and is highly susceptible to flooding. Northeastern Bangladesh experiences frequent extreme rainfall due to the orographic uplift caused by the surrounding hilly regions. Bangladesh experiences a tropical monsoon climate, with four distinct seasons: pre-monsoon summer (March to May), monsoon (June to September), post-monsoon (October to November), and winter (December to February). The monsoon season is particularly significant, accounting for most annual rainfall. Additionally, the pre- and post-monsoon seasons can bring tropical cyclones, causing extreme precipitation and coastal flooding.

\subsection{Data}
We downscale low-resolution ($0.25^{\circ}$ to $1^{\circ}$) model-simulated data to high-resolution ($0.05^{\circ}$) rainfall fields that are comparable to the gridded daily rainfall fields from Climate Hazards Group InfraRed Precipitation with Station data (CHIRPS). CHIRPS is a quasi-global rainfall dataset incorporating rainfall estimates from rain gauges and satellite observations~\cite{funk2015climate}. Our downscaling model trains using low-resolution data from the European Centre for Medium-Range Weather Forecasts (ECMWF) Reanalysis (ERA5), which is available at $0.25^{\circ}$ resolution~\cite{hersbach2020era5}. 

Due to inherent biases in ERA5 and CHIRPS, their rainfall fields do not align daily, presenting a challenge in training a downscaling function. A two-step downscaling process addresses this. Following the downscaling approach developed in SR24~\cite{saha2024statistical}, in the first step (GAN-1), ERA5 data ($0.25^{\circ}$) downscales to ERA5-Land resolution ($0.1^{\circ}$). ERA5-Land is a high-resolution replay of the land surface component of ERA5, providing finer spatial detail~\cite{munoz2021era5}. In the second step (GAN-2), upscaled CHIRPS rainfall fields become predictors for the original CHIRPS rainfall fields ($0.05^{\circ}$) as predictands. All the mentioned datasets were obtained for the years 1981 to 2019 and split into three groups: training (1981-1999), validation (2000-2009), and testing (2010-2019). Additionally, simulated outputs of seven models from the High-Resolution Model Intercomparison Project (HighResMIP) from the Coupled Model Intercomparison Project Phase 6 (CMIP6)~\cite{haarsma2016high} are obtained. The models are listed in Table~\ref{tab:highresmip}. This work defines 2000-2019 as the present climate and 2031-2050 as the future climate and assesses the changes between these periods.

\begin{table}[htpb]
    \centering
    \captionof{table}{List of HighResMIP models used in this study}
    \begin{tabular}{|l|c|}
    \hline
        \multicolumn{1}{|c|}{\textbf{Models}} & \textbf{Resolution} \\[1.1ex]
    \hline\hline
        FGOALS-f3-H & 25 Km \\
    \hline 
        HadGEM3-GC31-HH & 25 Km \\
    \hline
        HadGEM3-GC31-HM & 50 Km \\
    \hline
        HadGEM3-GC31-MM & 50 Km \\
    \hline
        HadGEM3-GC31-LL & 100 Km \\
    \hline
        MPI-ESM1-2-XR & 50 Km \\
    \hline
        MPI-ESM1-2-HR & 100 Km \\
    \hline
    \end{tabular}
    \label{tab:highresmip}
\end{table}

\begin{figure}[htpb]
\centering
\includegraphics[width=7cm]{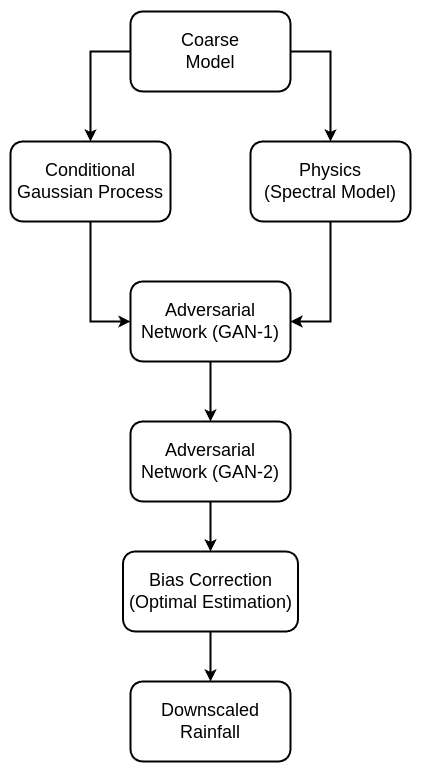}
\caption{A schematic representation of the downscaling method. Coarse-resolution model predictions are downscaled to the resolution of high-resolution observation with the help of a simple statistical model (conditional Gaussian process), simplified physics (spectral model), and two-step adversarial networks (GAN-1 and GAN-2). Optimal estimation is used to bias-correct the downscaled rainfall in the present climate.}
\label{fig:method}
\end{figure}

\subsection{Downscaling Method}
A schematic representation of our downscaling approach is presented in Figure~\ref{fig:method}. The approach consists of the following steps:
\begin{enumerate}
    \item A fast statistical model (conditional Gaussian process) to derive a ``first-guess" downscaled rainfall field.
    \item A simplified physics-based model (spectral method) to estimate orography-induced precipitation.
    \item an adversarial network (GAN-1) that combines these two fields and produces an improved downscaled rainfall field at $0.1^{\circ}$ resolution.
    \item Another adversarial network (GAN-2) that downscales the rainfall field to $0.05^{\circ}$ resolution
    \item An optimal estimation-based bias correction.
\end{enumerate}

Priming the downscaling model with statistics and physics-derived rainfall fields improves the physical consistency and alleviates data paucity issues. Two-step adversarial learning addresses the lack of correspondence between model and data and captures high-frequency details in the downscaled field. Biases between the model and data are corrected with optimal estimation to ensure that the downscaled rainfall captures the observed risk of extremes in the present climate.

\subsubsection{Conditional Gaussian Process}

We use an iterative conditional Gaussian process (CGP) regressor to produce a ``first-guess" downscaled rainfall field. First, we index the training pair of low- and high-resolution rainfall fields on a manifold. Let a low-resolution rainfall field be $L_q$, and $H_q$ its high-resolution counterpart to be generated. A nearest-neighbor search through the manifold of low-resolution fields yields the closest match to $L_q$, denoted as $L_k$. Also retrieved is the high-resolution counterpart $H_k$ associated with $L_k$. Next, $L_k$ and $H_k$ are iteratively improved until $L_q$ and $L_k$ converge, then resulting $H_k$ becomes the desired ``first-guess" downscaled field $H_q$.

\begin{equation}
\label{eqn:gaussianprocessdownscaling}
\begin{aligned}
  &L_k \leftarrow \text{Nearst Manifold Neigbhor Search}(L_q)\\
  [1]:\Rightarrow \; & e_k = L_q - L_k \\
  & H_{k} \leftarrow H_k + \alpha \mathbf{D}(e_k) \\
  & L_{k} \leftarrow \mathbf{U}(H_{k});\;\;\text{goto [1]}. 
\end{aligned}
\end{equation}

Here, $\mathbf{D}$ and $\mathbf{U}$ are downscaling and upscaling functions, respectively. This study uses averaging and pooling (subsampling) as the upscaling function and the following ensemble-approximated conditional Gaussian process regressor (Eqn.~\ref{eqn:cgp}) as the downscaling function. The rate $\alpha$ is set as a scaling constant.

\begin{equation}
\mathbf{D}(e_k) = {\mathrm{C}_{HL} \mathrm{C}_{LL}^{-1} e_k},
\label{eqn:cgp}
\end{equation}

where $\mathrm{C}_{LL}$ is the sample conditional covariance of low-resolution fields in the training manifold, and $\mathrm{C}_{HL}$ is the cross-covariance between low- and high-resolution fields in the training manifold. To overcome dimensionality issues, ensemble-based reduced-rank square-root methods are employed~\cite{ravela2007fast,ravela2010realtime,trautner2020informative}.

\subsubsection{Orographic Precipitation Estimation}
We use a linear spectral method to estimate the orographic component of precipitation~\cite{smith2004linear}.
\begin{equation}
\label{eqn:spectralmodel}
   \hat{P}(k,l) = \frac{C_w\;\iota\sigma\;\hat{h}(k,l)}{(1+\iota\sigma \tau_c)(1+\iota\sigma \tau_h)(1-\iota m(k,l)H_w)},
\end{equation}
where $k$ and $l$ are horizontal wavenumbers, $\iota=\sqrt{-1}$, $\hat{P}(k,l)$ is the orographic precipitation in the frequency domain, $\hat{h}(k,l)$ is the terrain elevation in the frequency domain, $u$ and $v$ are zonal and meridional components of wind, $\sigma=uk+vl$ is the corresponding intrinsic frequency~\cite{smith2004linear}, $C_w$ is the thermodynamic uplift sensitivity factor, a coefficient relating condensation rate to vertical motion (Eqn.~\ref{eqn:cw}), $\tau_c$ is the time constant for conversion from cloud water to hydrometeors, $\tau_h$ is the time constant for hydrometeor fallout, $m\doteq m(k,l)$ is the vertical wavenumber (Eqn.~\ref{eqn:vertical_wavenumber}), and $H_w$ is the depth of moist layer penetrated by vertical wind (Eqn.~\ref{eqn:depthmoistlayer}).
\begin{equation}
    C_w = \frac{1}{R_v} \frac{e_s(T)}{T} \frac{\Gamma_m}{\gamma},
\label{eqn:cw}
\end{equation}
where, $R_v$ is the gas constant of saturated air (461.5 J/Kg/K), $T$ is the air temperature, $e_s$ is the saturation vapor pressure, $\gamma$ is the environmental lapse rate, and $\Gamma_m$ is the moist adiabatic lapse rate. When the environmental lapse rate data is unavailable, we assume it to be 99\% of $\Gamma_m$, with a minimum of 6.5$^{\circ}$K/Km.
\begin{equation}
   e_{s} = e_{s0} \exp \left( \frac{L_v}{R_v} \left( \frac{1}{T_0} - \frac{1}{T} \right) \right),
\end{equation}
where, $L_v$ is latent heat of vaporization (2.26 $\times$ 10\textsuperscript{6} J/Kg) and $e_{s0}$ is the reference saturation vapor pressure at the reference temperature $T_0$. When $T_0$ is 273.16 K, $e_{s0}$ is 611 Pa.
\begin{equation}
    \Gamma_m = \frac{g}{C_p} \frac{\left(1+r_s\right)}{\left(1+r_s \frac{C_{pv}}{C_p}\right)} \frac{\left(1+\frac{L_v r_s}{R_d T}\right)}{\left(1+ \frac{L_v^2 r_s (1+ r_s \frac{R_v}{R_d})}{R_v T^2 (C_p + r_s C_{pv})}\right)},
\end{equation}
where $g$ is the gravitational acceleration, $R_d$ is the gas constant of dry air (287.04 J/Kg/K), $C_p$ is the specific heat of dry air (1003.5 J/Kg/K), $C_{pv}$ is the specific heat of the saturated air (1996 J/Kg/K), and $r_s$ is saturation mixing ratio.
\begin{equation}
    r_s = \frac{R_d}{R_v} \frac{e_s}{\left( p - e_s \right)},
\end{equation}
where $p$ is the surface pressure. When surface pressure data is unavailable, we assume it to be 1013.25 hPa.
\begin{equation}
    H_w = \frac{R_vT^2}{L_v\gamma}.
\label{eqn:depthmoistlayer}
\end{equation}
\begin{equation}
    m(k,l) = \sqrt{\frac{N_m^2 - \sigma^2}{\sigma^2}(k^2+l^2)}. sgn(\sigma),
    \label{eqn:vertical_wavenumber}
\end{equation}
where, $N_m^2$ is moist static stability, defined as
\begin{equation}
    N_m^2 = \frac{g}{T}(\gamma - \Gamma_m).
\end{equation}

\begin{figure*}[htpb]
\centering
\includegraphics[width=13cm]{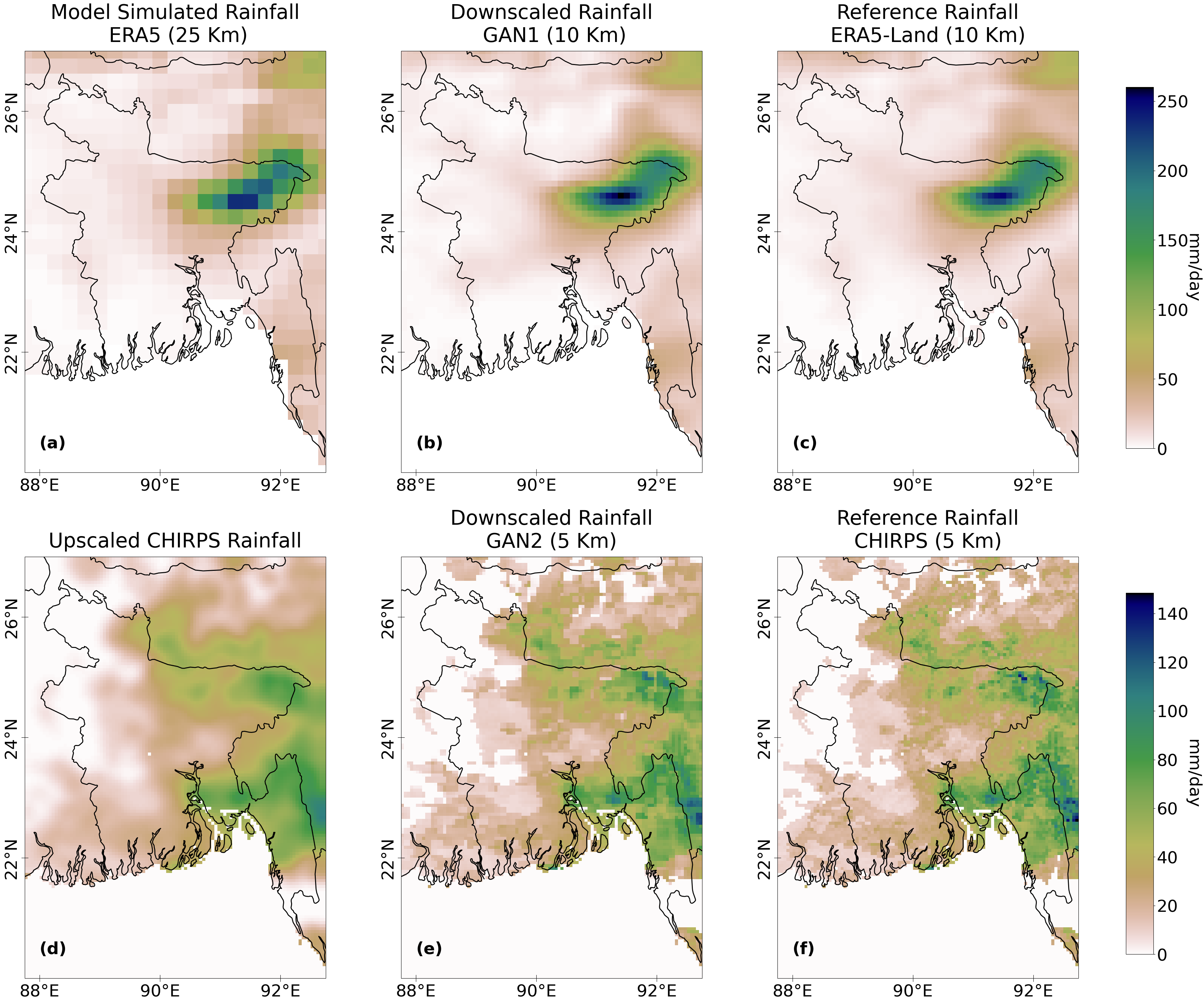}
\caption{A qualitative comparison of coarse, downscaled, and fine-resolution reference rainfall fields. The top row features an extreme event from ERA5/ERA5-Land and corresponding downscaled rainfall from GAN-1. The bottom row includes an extreme event from CHIRPS and corresponding downscaled rainfall from GAN-2. The panels are as follows: (a) Low-resolution ($0.25^{\circ}$) rainfall from ERA5, (b) downscaled rainfall from GAN-1 at resolution $0.1^{\circ}$,\newline (c) reference high-resolution ($0.1^{\circ}$) rainfall from ERA5-Land, (e) rainfall from CHIRPS that has been upscaled,\newline (f) downscaled rainfall from GAN-2 at resolution $0.05^{\circ}$, (g) original high-resolution ($0.05^{\circ}$) reference rainfall from CHIRPS. Comparison of the middle and right columns highlights the effectiveness of our downscaling model.}
\label{fig:downscalingmodeleval}
\end{figure*}

\subsubsection{Adversarial Learning}
Adversarial learning employs two competing networks, a Generator ($\mathcal{G}$) and a Discriminator ($\mathcal{D}$)~\cite{goodfellow2014generative}. The Generator $\mathcal{G}$($L;\alpha_{\mathcal{G}})$ maps the input low-resolution ($L$) rainfall to a super-resolution reconstruction using a deep convolutional and upsampling network. The Discriminator $\mathcal{D}$($L;\alpha_{\mathcal{D}})$, repeatedly fed high-resolution reference rainfall fields ($H$) and the super-resolution generator output, learns to tell them apart. The Generator tries to increase $\log \mathcal{D}(\mathcal{G}(L))$, while the Discriminator tries to reduce it in a minimax optimization game. The following loss functions apply to the discriminator and generator, respectively~\cite{jolicoeur2018relativistic}.
\begin{alignat*}{3}
    \mathcal{L}_\mathcal{D} (L, H) &= &&- \mathbb{E}_H &&\biggl[ \log \Bigl[ \mathcal{D} \bigl(H\bigr) - \mathbb{E}_L \bigl[ \mathcal{D} \bigl( \mathcal{G} \left( L \right) \bigr) \bigr] \Bigr] \biggr]\\
    & &&+ \mathbb{E}_L &&\biggl[ \log \Bigl[ \mathcal{D} \bigl(\mathcal{G}(L) \bigr) - \mathbb{E}_H \bigl[ \mathcal{D} \bigl(H\bigr) \bigr] \Bigr] \biggr]
    \addtocounter{equation}{1}\tag{\theequation}
\end{alignat*}
\begin{alignat*}{3}
    \mathcal{L}_\mathcal{G} (L, H) &= && &&\mathcal{L}_\mathcal{G}^{adv.} + \lambda \mathcal{L}_\mathcal{G}^{\ell_1} \\
    &= &&- &&\mathbb{E}_L \biggl[ \log \Bigl[ \mathcal{D}\bigl(\mathcal{G}(L)\bigr) - \mathbb{E}_H \bigl[ \mathcal{D}\bigl(H\bigr) \bigr] \Bigr] \biggr]\\
    & &&+ &&\lambda \ell_1\bigl[\mathcal{G}(L)\bigr]
    \addtocounter{equation}{1}\tag{\theequation}
\end{alignat*}
where, $\mathbb{E}$ is the expectation operator, $\ell_1$ is the total pixel-wise loss, and $\lambda \in \mathbb{R}$ is a regularization factor tunable as a hyperparameter. A point of note is that pixel-wise loss is necessary to maintain correspondence with the input low-resolution field. However, as did other papers in the literature, we noticed that doing so may also introduce blurriness in the high-resolution output~\cite{ledig2017photo, fritsche2019frequency}. To remedy the issue, the reconstructed field is segregated into low- and high-frequency components with a low-pass Gaussian filter before calculating the losses. Adversarial losses ($\mathcal{L}_\mathcal{D}$ and $\mathcal{L}_\mathcal{G}^{adv.}$) now apply in the high-frequency component, and pixel-wise loss($\ell_1$) applies in the low-frequency component~\cite{fritsche2019frequency}. Following SR24, a densely connected Residual in Residual Dense Blocks (RRDB) network is the generator, and a VGG-style deep convolutional network is the discriminator~\cite{saha2024statistical}.

\subsubsection{Bias Correction}

We employ the peak-over-threshold method to identify extreme rainfall and determine their return periods. The return period, $R$, is given by,
\begin{equation}
    R = \frac{1}{E_p \lambda},
\end{equation}
where, $E_p$ represents the exceedance probability, and $\lambda$ is the frequency of extremes. A two-parameter Generalized Pareto distribution is fitted to the estimated return periods, whose probability density function is given by,
\begin{equation}
      f (r| k, \sigma) = \frac{1}{\sigma} \left( 1 + \frac{kr}{\sigma} \right) ^ {-1-\frac{1}{k}}.
\end{equation}
Unlike SR24, the Pareto distribution fits only to the upper tail, and a kernel density estimate fits the remaining data.  Bootstrapping generates multiple instances to account for uncertainty in the risk curves, and optimal estimation corrects bias between the downscaled and observed risk curves~\cite{rodgers1976retrieval}.
\begin{equation}
\label{eqn:optimal_estimation}
    y_{d}^* = y_{d} + \mathrm{C}_{dd} \left( \mathrm{C}_{dd} + \mathrm{C}_{oo} \right)^{-1}(y_{o} - y_{d}).
\end{equation}
Here, $y_{o}$, $y_{d}$, and $y_{d}^*$ denote the observed, downscaled, and bias-corrected risk curves, respectively, and $\mathrm{C}$ is the covariance operator with $\mathrm{C}_{dd}$ the covariance of downscaled risk curve, and $\mathrm{C}_{oo}$ of the observed risk curve.  The covariance is calculated for the ensemble of risk curves generated by bootstrapping. Data from the validation period estimates the bias correction applied to the testing period. Finally, the corrected risk curves are back-projected onto rainfall fields via quantile mapping.

\subsubsection{Application on HighResMIP Models}
We apply the downscaling model trained on the ERA5 reanalysis model on seven HighResMIP climate model outcomes. Some models are coarser than ERA5 (see Table~\ref{tab:highresmip}). A bicubic interpolation brings them to the same resolution as ERA5 before applying the conditional Gaussian process. The same downscaling function is applied to present and future climate projections, assuming that the downscaling function remains unchanged in the warming scenario.

\section{Results}
\label{section:results}

Figure~\ref{fig:downscalingmodeleval} presents a qualitative evaluation of the downscaling model against reference rainfall fields (``truth"). Figure~\ref{fig:downscalingmodeleval}a shows an extreme rainfall event from ERA5 at a resolution of $0.25^{\circ}$, which is downscaled to $0.1^{\circ}$ by GAN-1 in Figure~\ref{fig:downscalingmodeleval}b, and compared with the corresponding rainfall field produced by ERA5-Land ($0.1^{\circ}$) in Figure~\ref{fig:downscalingmodeleval}c. Similarly, a rainfall event from the CHIRPS dataset ($0.05^{\circ}$), shown in Figure~\ref{fig:downscalingmodeleval}f, is upscaled in Figure~\ref{fig:downscalingmodeleval}d, and then downscaled to Figure~\ref{fig:downscalingmodeleval}e by GAN-2. Comparing the outcomes of both GAN-1 and GAN-2 with their respective references shows that our model sufficiently captures the spatial pattern and heterogeneity of the high-resolution rainfall.

\begin{figure}[htpb]
\centering
\includegraphics[width=7cm]{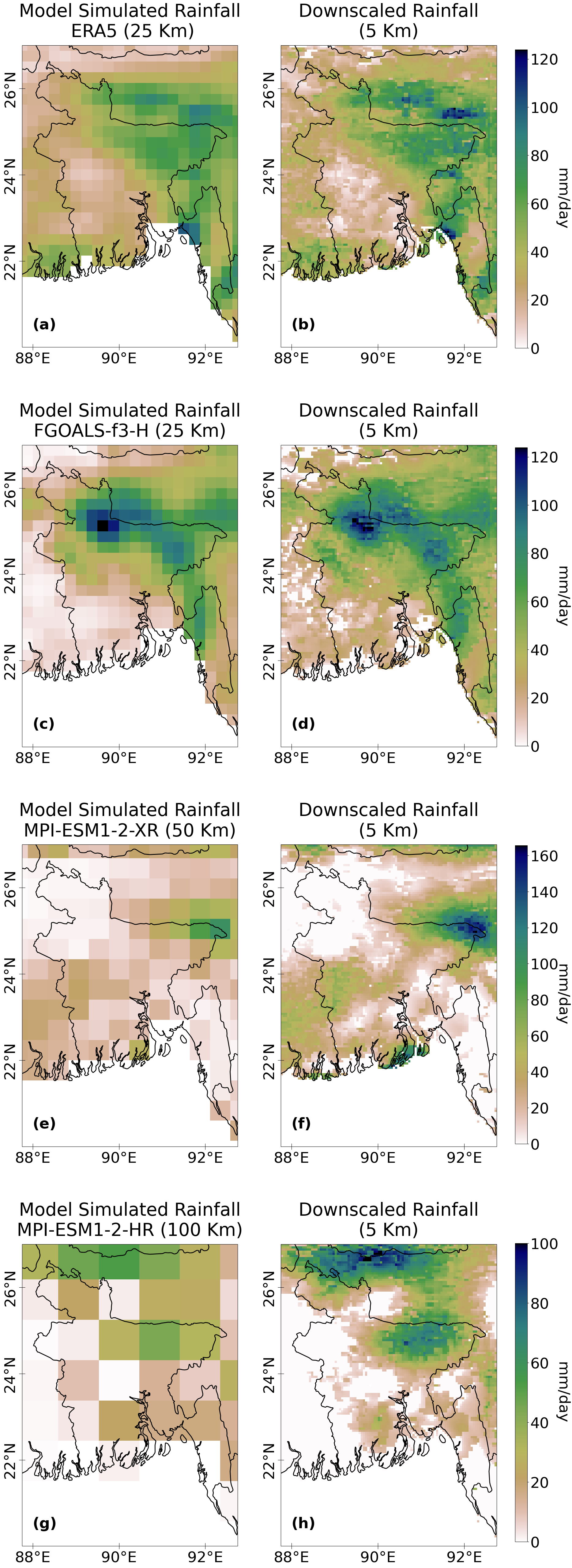}
\caption{Downscaling various models with different resolutions. The left column shows the coarse rainfall fields simulated by the models, while the right column shows the corresponding downscaled rainfall fields. The top row features rainfall fields from the reanalysis model ERA5 (resolution $0.25^{\circ}$). The subsequent three rows present rainfall fields from three HighResMIP models: FGOALS-f3-H (resolution $0.25^{\circ}$), MPI-ESM2-XR (resolution $0.5^{\circ}$), and MPI-ESM2-HR (resolution $1^{\circ}$). Although the downscaling model was trained using $0.25^{\circ}$ resolution model, it can effectively downscale coarser $1^{\circ}$ models.}
\label{fig:downscalingapplication}
\end{figure}

Figure~\ref{fig:downscalingapplication} showcases downscaling applied to different coarse-resolution climate model outputs of varying resolutions. A direct comparison with a reference rainfall field is impossible here due to a lack of correspondence between the model and observation on a daily scale. However, we observe that the downscaled rainfall closely resembles the high-frequency data of the CHIRPS rainfall, indicating the effectiveness of our approach. Even though the downscaling model was trained against $0.25^{\circ}$ resolution data, applying it to coarser resolution models has been reasonably successful.

\begin{figure*}[htpb]
\centering
\includegraphics[width=15cm]{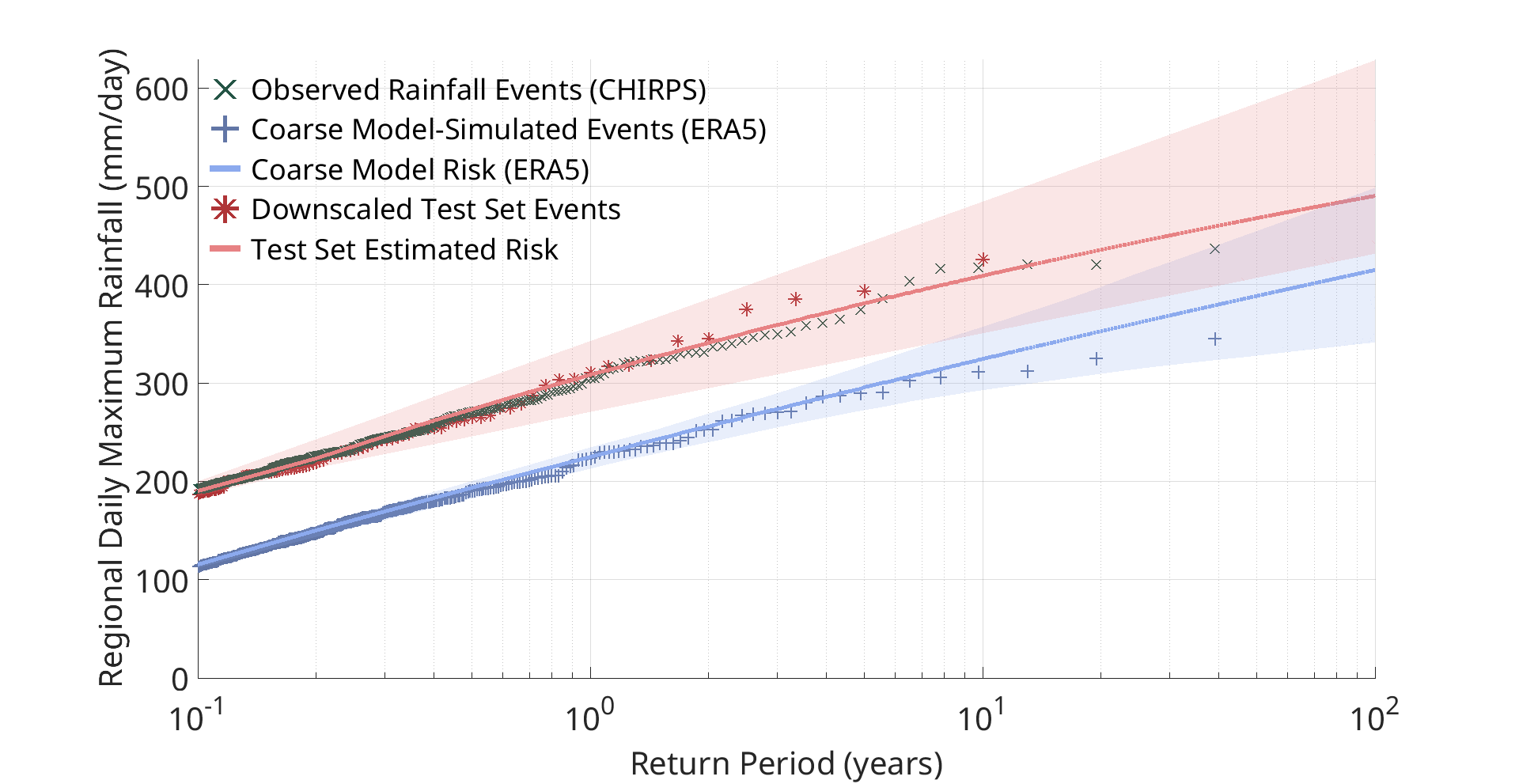}
\caption{Comparison of extreme rainfall risks as captured by a coarse model (ERA5), high-resolution observations (CHIRPS), and downscaled rainfall data in the present climate. The figure displays daily regional rainfall maxima plotted against their return periods, with individual events from ERA5 shown as blue plus signs, CHIRPS as green crosses, and downscaled rainfall as red asterisks. Solid lines represent Generalized Pareto fits through these events, extended to a 100-year return period, with the shaded area indicating uncertainty. This figure illustrates that coarse models tend to underestimate extreme rainfall risks, but downscaling can better capture the observed risk in the present climate.}
\label{fig:riskpresent}
\end{figure*}

Figure~\ref{fig:riskpresent} assesses the ability of downscaled rainfall data to reflect observed risk in the present climate. This figure presents the daily maxima of extreme events from CHIRPS, ERA5, and downscaled ERA5 rainfall against their return periods. Two-parameter Generalized Pareto distributions fitted through them are presented with solid lines, with uncertainty around these lines estimated through bootstrapping. It is evident that the coarse-resolution ERA5 model significantly underestimates extreme risk, highlighting the necessity for downscaling. The results indicate that downscaled rainfall effectively captures the observed risk. Given that the downscaling model accurately represents both the spatial patterns and the risk of high-resolution rainfall in the present climate, we apply it to both present and future climate data from the HighResMIP models, with the assumption that the downscaling function is invariant to the climate change.

\begin{figure*}[htpb]
\centering
\includegraphics[width=15cm]{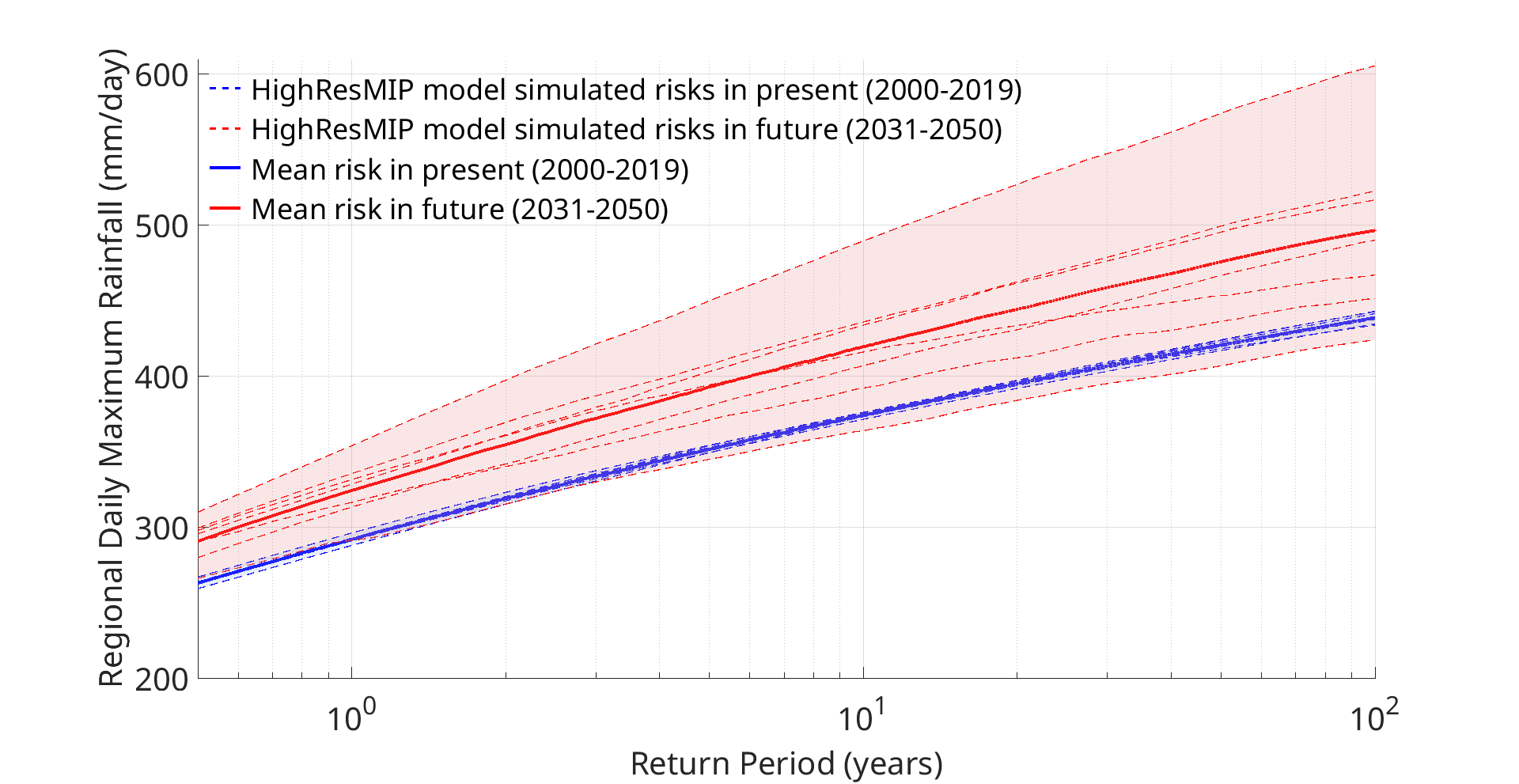}
\caption{Comparison of extreme rainfall risks captured by downscaled HighResMIP models for the present climate (2000-2019, shown in blue) and the future climate (2031-2050, shown in red). Dotted lines represent Generalized Pareto fits for daily regional rainfall maxima from individual models, plotted against their return periods. Solid lines denote the mean. The shaded area represents the inter-model spread. This figure displays the expected increase in the extreme rainfall risk from the present to the future climate and the model uncertainty around it.}
\label{fig:riskfuture}
\end{figure*}

Figure~\ref{fig:riskfuture} compares the extreme rainfall risks captured by downscaled rainfall from seven HighResMIP models for both the present and future climates. Since present climate data is individually bias-corrected against observations for each model, the variation between models is minimal. However, there is less consensus among the models in the future climate, as indicated by a more extensive inter-model spread. Most models project an increase in extreme risk for the future climate, with an expected rise of approximately 50 mm/day in the return level of a 100-year return period.

\begin{figure*}[htpb]
\centering
\includegraphics[width=13cm]{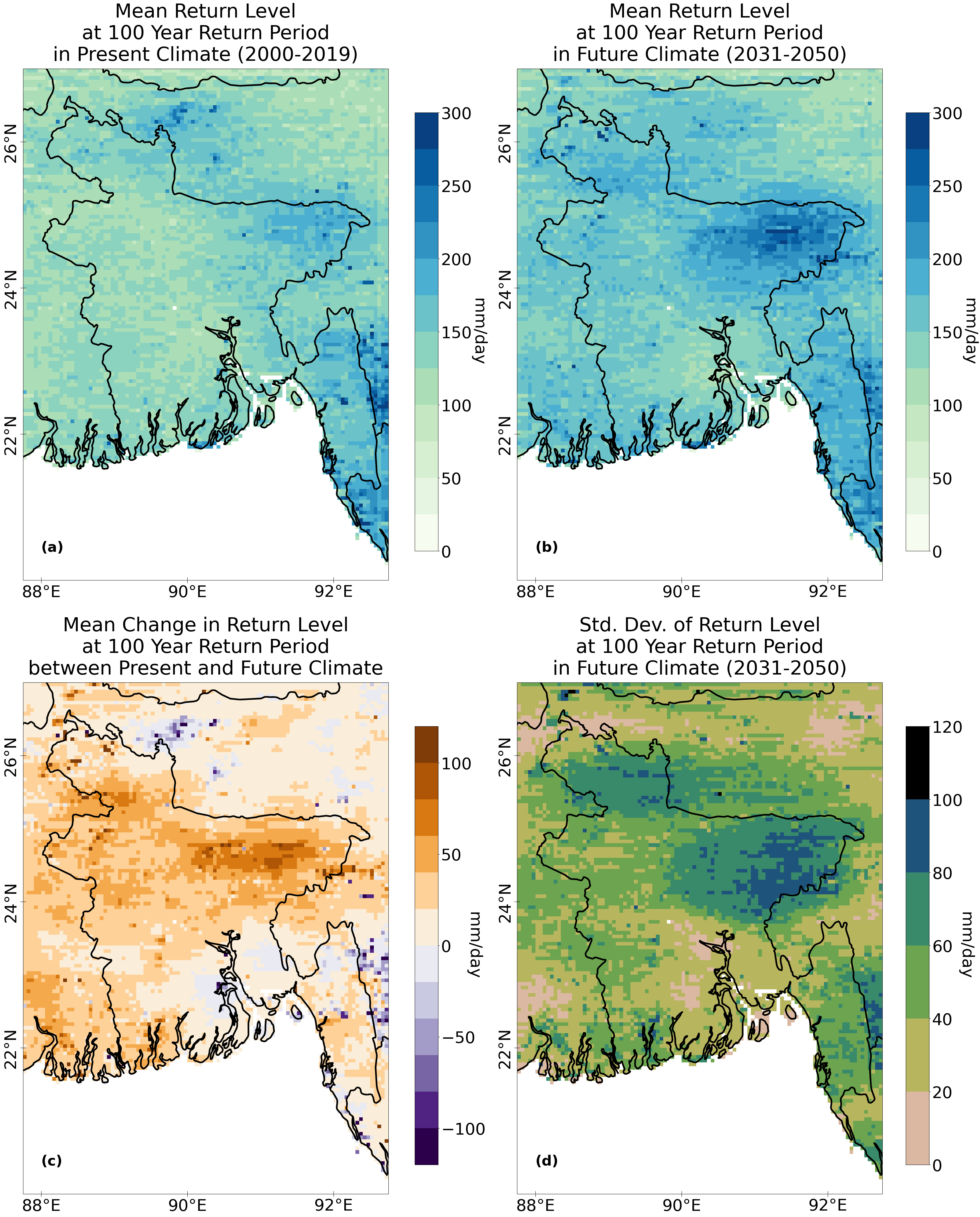}
\caption{Spatial distribution of extreme rainfall risk captured by downscaled HighResMIP models. Panel (a) shows the mean return level at a 100-year return period for the present climate (2000-2019), while panel (b) illustrates the same for the future climate (2031-2050). Panel (c) displays the difference between the future and present climates. Panel (d) depicts the inter-model standard deviation for the future climate shown in (b).}
\label{fig:riskfuturespatial}
\end{figure*}

Figure~\ref{fig:riskfuturespatial} illustrates the spatial distribution of extreme rainfall risk captured by downscaled HighResMIP models. Comparing the mean return level for a 100-year return period between the present and future climates helps identify areas where the risk of extremes is expected to increase and the extent of this increase. As shown in Figure~\ref{fig:riskfuturespatial}c, the maximum rise in extreme risk will likely occur in Northeast Bangladesh, a region already susceptible to extremes due to its surrounding topography. However, model projections have substantial variability, as evident in Figure~\ref{fig:riskfuturespatial}d.

\section{Discussion}
\label{section:discussions}
This study employs a rapid rainfall downscaling method to assess the risk of extremes in Bangladesh in a warming climate. Bangladesh's geographic location and socio-economic conditions make it highly susceptible to climate-related risks. Coarse climate models are inadequate for capturing this risk, which motivates downscaling. Dynamic downscaling techniques are computationally expensive and slow, making them impractical for uncertainty quantification. Data-driven techniques also often face challenges, such as limited data, lack of direct correspondence between model predictions and observations, and issues with physical consistency. SR24 developed a novel downscaling method, which integrates statistics, physics, and machine learning, to address these challenges~\cite{saha2024statistical}. Priming the machine learning model with Statistics reduces the need for extensive data augmentation, and incorporating physics improves the physical consistency of the results. Even though each model component has limitations, they complement each other, collectively improving the approach's efficacy. While SR24 focused on urban-scale extremes in the current climate, this study extends the method to a national scale for both present and future climates. In particular, introducing frequency separation during the ML model training refines the model's capability to capture heterogeneous spatial patterns. We also improve risk assessment with an improved distribution parameterization using bootstrapping for uncertainty assessment. Our approach effectively evaluates model uncertainty by training the model once and applying it across multiple climate models.

Downscaling seven HighResMIP models shows that Bangladesh's risk of extreme events will increase. This finding aligns with the global consensus on climate change. However, climate models do not have consensus on a regional scale~\cite{saha2019can}, and the results show substantial variability across models. There are also uncertainties associated with the warming scenario. The HighResMIP models provide projections based on a single scenario (SSP5-8.5) reported in this manuscript. Future research could involve downscaling CMIP6 models with multiple scenario projections to understand scenario uncertainty better. 

One fundamental assumption of our approach is that the downscaling function is invariant to climate change. This assumption enables us to train the downscaling function using present climate data and apply it to future scenarios. It is unclear if this nominal assumption holds. Training the downscaling model on low- and high-resolution simulations covering current and future climates would help address this question. However, such simulations are not readily available due to their high computational costs. Another area for future research is investigating the risk of cascading extreme events, such as extreme rainfall accompanied or followed by cyclones, heat stress, or floods. Our downscaling approach can be extended to other variables of interest (such as wind, temperature, and inundation) to assess these risks in future climate within an integrated framework. Finally, we are interested in recovering continuous scaling laws for model fields (scale spaces~\cite{ravela2003multi}), using geometry-coupled random field models~\cite{ravela2014spatial,ravela2015dynamic} with neural dynamics~\cite{trautner2020informative}.

\section*{Acknowledgment}
The authors acknowledge support from the Jameel Observatory CREWSNet project, Eric and Wendy Schmidt, by recommendation of Schmidt Futures as part of its Virtual Earth System Research Institute (VESRI), the MIT Climate Grand Challenges program, and Liberty Mutual. The authors thank Dr. Jiangchao Qiu for discussions and return period calculations.

\section*{Data Availability Statement}
All datasets used in this study are publicly available. CHIRPS global daily rainfall dataset is available from the Climate Hazards Center, University of California, Santa Barbara, USA (\url{https://data.chc.ucsb.edu/products/CHIRPS-2.0/}). ERA5 and ERA5-land reanalysis data are downloaded from Copernicus Climate Change Service (C3S) Climate Data Store (CDS) (\url{https://cds.climate.copernicus.eu}). HighResMIP data is obtained from the Earth System Grid Federation (ESGF) nodes (\url{https://aims2.llnl.gov}). The topographic elevation data obtained by Shuttle Radar Topography Mission (SRTM) at 90 m horizontal resolution is acquired from OpenTopography (\url{https://portal.opentopography.org/raster?opentopoID=OTSRTM.042013.4326.1}). Downscaled extreme rainfall fields from ERA5 and HighResMIP models produced in this study are available at \url{https://doi.org/10.5281/zenodo.13351177}.

\bibliographystyle{unsrtnat}
\bibliography{references}

\begin{thebibliography}{59}
\providecommand{\natexlab}[1]{#1}
\providecommand{\url}[1]{\texttt{#1}}
\expandafter\ifx\csname urlstyle\endcsname\relax
  \providecommand{\doi}[1]{doi: #1}\else
  \providecommand{\doi}{doi: \begingroup \urlstyle{rm}\Url}\fi

\bibitem[Eckstein et~al.(2021)Eckstein, K{\"u}nzel, and Sch{\"a}fer]{eckstein2021global}
David Eckstein, Vera K{\"u}nzel, and Laura Sch{\"a}fer.
\newblock \emph{The global climate risk index 2021}.
\newblock Bonn: Germanwatch, 2021.

\bibitem[Schulthess et~al.(2018)Schulthess, Bauer, Wedi, Fuhrer, Hoefler, and Sch{\"a}r]{schulthess2018reflecting}
Thomas~C Schulthess, Peter Bauer, Nils Wedi, Oliver Fuhrer, Torsten Hoefler, and Christoph Sch{\"a}r.
\newblock Reflecting on the goal and baseline for exascale computing: a roadmap based on weather and climate simulations.
\newblock \emph{Computing in Science \& Engineering}, 21\penalty0 (1):\penalty0 30--41, 2018.

\bibitem[Prein et~al.(2020)Prein, Liu, Ikeda, Bullock, Rasmussen, Holland, and Clark]{prein2020simulating}
Andreas~F Prein, Changhai Liu, Kyoko Ikeda, Randy Bullock, Roy~M Rasmussen, Greg~J Holland, and Martyn Clark.
\newblock Simulating north american mesoscale convective systems with a convection-permitting climate model.
\newblock \emph{Climate Dynamics}, 55:\penalty0 95--110, 2020.

\bibitem[Saha and Ravela(2024)]{saha2024statistical}
Anamitra Saha and Sai Ravela.
\newblock Statistical-physical adversarial learning from data and models for downscaling rainfall extremes.
\newblock \emph{Journal of Advances in Modeling Earth Systems}, 16\penalty0 (6):\penalty0 e2023MS003860, 2024.

\bibitem[S{\o}rland et~al.(2021)S{\o}rland, Brogli, Pothapakula, Russo, Van~de Walle, Ahrens, Anders, Bucchignani, Davin, Demory, et~al.]{sorland2021cosmo}
Silje~Lund S{\o}rland, Roman Brogli, Praveen~Kumar Pothapakula, Emmanuele Russo, Jonas Van~de Walle, Bodo Ahrens, Ivonne Anders, Edoardo Bucchignani, Edouard~L Davin, Marie-Estelle Demory, et~al.
\newblock Cosmo-clm regional climate simulations in the coordinated regional climate downscaling experiment (cordex) framework: a review.
\newblock \emph{Geoscientific Model Development}, 14\penalty0 (8):\penalty0 5125--5154, 2021.

\bibitem[Katragkou et~al.(2015)Katragkou, Garc{\'\i}a-D{\'\i}ez, Vautard, Sobolowski, Zanis, Alexandri, Cardoso, Colette, Fernandez, Gobiet, et~al.]{katragkou2015regional}
Eleni Katragkou, Markel Garc{\'\i}a-D{\'\i}ez, Robert Vautard, S~Sobolowski, Prodromos Zanis, G~Alexandri, Rita~M Cardoso, Augustin Colette, J~Fernandez, A~Gobiet, et~al.
\newblock Regional climate hindcast simulations within euro-cordex: evaluation of a wrf multi-physics ensemble.
\newblock \emph{Geoscientific Model Development}, 8\penalty0 (3):\penalty0 603--618, 2015.

\bibitem[Sakaguchi et~al.(2023)Sakaguchi, Leung, Zarzycki, Jang, McGinnis, Harrop, Skamarock, Gettelman, Zhao, Gutowski, et~al.]{sakaguchi2023technical}
Koichi Sakaguchi, L~Ruby Leung, Colin~M Zarzycki, Jihyeon Jang, Seth McGinnis, Bryce~E Harrop, William~C Skamarock, Andrew Gettelman, Chun Zhao, William~J Gutowski, et~al.
\newblock Technical descriptions of the experimental dynamical downscaling simulations over north america by the cam--mpas variable-resolution model.
\newblock \emph{Geoscientific Model Development}, 16\penalty0 (10):\penalty0 3029--3081, 2023.

\bibitem[Roe(2005)]{roe2005orographic}
Gerard~H Roe.
\newblock Orographic precipitation.
\newblock \emph{Annu. Rev. Earth Planet. Sci.}, 33:\penalty0 645--671, 2005.

\bibitem[Smith and Barstad(2004)]{smith2004linear}
Ronald~B Smith and Idar Barstad.
\newblock A linear theory of orographic precipitation.
\newblock \emph{Journal of the Atmospheric Sciences}, 61\penalty0 (12):\penalty0 1377--1391, 2004.

\bibitem[Paeth et~al.(2017)Paeth, Pollinger, M{\"a}chel, Figura, Wahl, Ohlwein, and Hense]{paeth2017efficient}
Heiko Paeth, Felix Pollinger, Hermann M{\"a}chel, Clarissa Figura, Sabrina Wahl, Christian Ohlwein, and Andreas Hense.
\newblock An efficient model approach for very high resolution orographic precipitation.
\newblock \emph{Quarterly Journal of the Royal Meteorological Society}, 143\penalty0 (706):\penalty0 2221--2234, 2017.

\bibitem[Wood et~al.(2002)Wood, Maurer, Kumar, and Lettenmaier]{wood2002long}
Andrew~W Wood, Edwin~P Maurer, Arun Kumar, and Dennis~P Lettenmaier.
\newblock Long-range experimental hydrologic forecasting for the eastern united states.
\newblock \emph{Journal of Geophysical Research: Atmospheres}, 107\penalty0 (D20):\penalty0 ACL--6, 2002.

\bibitem[Sachindra et~al.(2014)Sachindra, Huang, Barton, and Perera]{sachindra2014multi}
DA~Sachindra, Fuchun Huang, AF~Barton, and BJC Perera.
\newblock Multi-model ensemble approach for statistically downscaling general circulation model outputs to precipitation.
\newblock \emph{Quarterly Journal of the Royal Meteorological Society}, 140\penalty0 (681):\penalty0 1161--1178, 2014.

\bibitem[Beecham et~al.(2014)Beecham, Rashid, and Chowdhury]{beecham2014statistical}
Simon Beecham, Mamunur Rashid, and Rezaul~K Chowdhury.
\newblock Statistical downscaling of multi-site daily rainfall in a south australian catchment using a generalized linear model.
\newblock \emph{International journal of climatology}, 34\penalty0 (14), 2014.

\bibitem[Lall et~al.(1996)Lall, Rajagopalan, and Tarboton]{lall1996nonparametric}
Upmanu Lall, Balaji Rajagopalan, and David~G Tarboton.
\newblock A nonparametric wet/dry spell model for resampling daily precipitation.
\newblock \emph{Water resources research}, 32\penalty0 (9):\penalty0 2803--2823, 1996.

\bibitem[Kannan and Ghosh(2013)]{kannanNonparametricKernelRegression2013}
S.~Kannan and Subimal Ghosh.
\newblock A nonparametric kernel regression model for downscaling multisite daily precipitation in the {{Mahanadi}} basin.
\newblock \emph{Water Resources Research}, 49\penalty0 (3):\penalty0 1360--1385, 2013.
\newblock ISSN 1944-7973.
\newblock \doi{10.1002/wrcr.20118}.

\bibitem[Mehrotra and Sharma(2005)]{mehrotra2005nonparametric}
R~Mehrotra and Ashish Sharma.
\newblock A nonparametric nonhomogeneous hidden markov model for downscaling of multisite daily rainfall occurrences.
\newblock \emph{Journal of Geophysical Research: Atmospheres}, 110\penalty0 (D16), 2005.

\bibitem[Zhang and Yan(2015)]{zhang2015new}
Xianliang Zhang and Xiaodong Yan.
\newblock A new statistical precipitation downscaling method with bayesian model averaging: a case study in china.
\newblock \emph{Climate Dynamics}, 45\penalty0 (9):\penalty0 2541--2555, 2015.

\bibitem[He et~al.(2016)He, Chaney, Schleiss, and Sheffield]{heSpatialDownscalingPrecipitation2016}
Xiaogang He, Nathaniel~W. Chaney, Marc Schleiss, and Justin Sheffield.
\newblock Spatial downscaling of precipitation using adaptable random forests.
\newblock \emph{Water Resources Research}, 52\penalty0 (10):\penalty0 8217--8237, 2016.
\newblock ISSN 1944-7973.
\newblock \doi{10.1002/2016WR019034}.

\bibitem[Tripathi et~al.(2006)Tripathi, Srinivas, and Nanjundiah]{tripathiDownscalingPrecipitationClimate2006}
Shivam Tripathi, V.~V. Srinivas, and Ravi~S. Nanjundiah.
\newblock Downscaling of precipitation for climate change scenarios: {{A}} support vector machine approach.
\newblock \emph{Journal of Hydrology}, 330\penalty0 (3):\penalty0 621--640, November 2006.
\newblock ISSN 0022-1694.
\newblock \doi{10.1016/j.jhydrol.2006.04.030}.

\bibitem[George and P(2023)]{george2023multi}
Jose George and Athira P.
\newblock A multi-stage stochastic approach for statistical downscaling of rainfall.
\newblock \emph{Water Resources Management}, 37\penalty0 (14):\penalty0 5477--5492, 2023.

\bibitem[Sachindra and Kanae(2019)]{sachindraMachineLearningDownscaling2019}
D.~A. Sachindra and S.~Kanae.
\newblock Machine learning for downscaling: The use of parallel multiple populations in genetic programming.
\newblock \emph{Stochastic Environmental Research and Risk Assessment}, 33\penalty0 (8):\penalty0 1497--1533, September 2019.
\newblock ISSN 1436-3259.
\newblock \doi{10.1007/s00477-019-01721-y}.

\bibitem[Wang et~al.(2020)Wang, Huang, Liu, Men, Guo, Miao, Jiao, Wang, Shoaib, and Xia]{wangSequencebasedStatisticalDownscaling2020}
Qingrui Wang, Jing Huang, Ruimin Liu, Cong Men, Lijia Guo, Yuexi Miao, Lijun Jiao, Yifan Wang, Muhammad Shoaib, and Xinghui Xia.
\newblock Sequence-based statistical downscaling and its application to hydrologic simulations based on machine learning and big data.
\newblock \emph{Journal of Hydrology}, 586:\penalty0 124875, July 2020.
\newblock ISSN 0022-1694.
\newblock \doi{10.1016/j.jhydrol.2020.124875}.

\bibitem[Tran~Anh et~al.(2019)Tran~Anh, Van, Dang, and Hoang]{tran2019downscaling}
Duong Tran~Anh, Song~P Van, Thanh~D Dang, and Long~P Hoang.
\newblock Downscaling rainfall using deep learning long short-term memory and feedforward neural network.
\newblock \emph{International Journal of Climatology}, 39\penalty0 (10):\penalty0 4170--4188, 2019.

\bibitem[Vandal et~al.(2019)Vandal, Kodra, and Ganguly]{vandalIntercomparisonMachineLearning2019}
Thomas Vandal, Evan Kodra, and Auroop~R. Ganguly.
\newblock Intercomparison of machine learning methods for statistical downscaling: The case of daily and extreme precipitation.
\newblock \emph{Theoretical and Applied Climatology}, 137\penalty0 (1):\penalty0 557--570, July 2019.
\newblock ISSN 1434-4483.
\newblock \doi{10.1007/s00704-018-2613-3}.

\bibitem[Reddy et~al.(2023)Reddy, Matear, Taylor, Thatcher, and Grose]{reddy2023precipitation}
P~Jyoteeshkumar Reddy, Richard Matear, John Taylor, Marcus Thatcher, and Michael Grose.
\newblock A precipitation downscaling method using a super-resolution deconvolution neural network with step orography.
\newblock \emph{Environmental Data Science}, 2:\penalty0 e17, 2023.

\bibitem[Vandal et~al.(2017)Vandal, Kodra, Ganguly, Michaelis, Nemani, and Ganguly]{vandal2017deepsd}
Thomas Vandal, Evan Kodra, Sangram Ganguly, Andrew Michaelis, Ramakrishna Nemani, and Auroop~R Ganguly.
\newblock Deepsd: Generating high resolution climate change projections through single image super-resolution.
\newblock In \emph{Proceedings of the 23rd acm sigkdd international conference on knowledge discovery and data mining}, pages 1663--1672, 2017.

\bibitem[Dong et~al.(2015)Dong, Loy, He, and Tang]{dong2015image}
Chao Dong, Chen~Change Loy, Kaiming He, and Xiaoou Tang.
\newblock Image super-resolution using deep convolutional networks.
\newblock \emph{IEEE transactions on pattern analysis and machine intelligence}, 38\penalty0 (2):\penalty0 295--307, 2015.

\bibitem[Kim et~al.(2016)Kim, Lee, and Lee]{kim2016accurate}
Jiwon Kim, Jung~Kwon Lee, and Kyoung~Mu Lee.
\newblock Accurate image super-resolution using very deep convolutional networks.
\newblock In \emph{Proceedings of the IEEE conference on computer vision and pattern recognition}, pages 1646--1654, 2016.

\bibitem[Lim et~al.(2017)Lim, Son, Kim, Nah, and Mu~Lee]{lim2017enhanced}
Bee Lim, Sanghyun Son, Heewon Kim, Seungjun Nah, and Kyoung Mu~Lee.
\newblock Enhanced deep residual networks for single image super-resolution.
\newblock In \emph{Proceedings of the IEEE conference on computer vision and pattern recognition workshops}, pages 136--144, 2017.

\bibitem[Lai et~al.(2017)Lai, Huang, Ahuja, and Yang]{lai2017deep}
Wei-Sheng Lai, Jia-Bin Huang, Narendra Ahuja, and Ming-Hsuan Yang.
\newblock Deep laplacian pyramid networks for fast and accurate super-resolution.
\newblock In \emph{Proceedings of the IEEE conference on computer vision and pattern recognition}, pages 624--632, 2017.

\bibitem[Li et~al.(2018)Li, Fang, Mei, and Zhang]{li2018multi}
Juncheng Li, Faming Fang, Kangfu Mei, and Guixu Zhang.
\newblock Multi-scale residual network for image super-resolution.
\newblock In \emph{Proceedings of the European conference on computer vision (ECCV)}, pages 517--532, 2018.

\bibitem[Ledig et~al.(2017)Ledig, Theis, Husz{\'a}r, Caballero, Cunningham, Acosta, Aitken, Tejani, Totz, Wang, et~al.]{ledig2017photo}
Christian Ledig, Lucas Theis, Ferenc Husz{\'a}r, Jose Caballero, Andrew Cunningham, Alejandro Acosta, Andrew Aitken, Alykhan Tejani, Johannes Totz, Zehan Wang, et~al.
\newblock Photo-realistic single image super-resolution using a generative adversarial network.
\newblock In \emph{Proceedings of the IEEE conference on computer vision and pattern recognition}, pages 4681--4690, 2017.

\bibitem[Wang et~al.(2018)Wang, Yu, Wu, Gu, Liu, Dong, Qiao, and Change~Loy]{wang2018esrgan}
Xintao Wang, Ke~Yu, Shixiang Wu, Jinjin Gu, Yihao Liu, Chao Dong, Yu~Qiao, and Chen Change~Loy.
\newblock Esrgan: Enhanced super-resolution generative adversarial networks.
\newblock In \emph{Proceedings of the European conference on computer vision (ECCV) workshops}, pages 1--16, 2018.

\bibitem[Hess et~al.(2022)Hess, Dr{\"u}ke, Petri, Strnad, and Boers]{hess2022physically}
Philipp Hess, Markus Dr{\"u}ke, Stefan Petri, Felix~M Strnad, and Niklas Boers.
\newblock Physically constrained generative adversarial networks for improving precipitation fields from earth system models.
\newblock \emph{Nature Machine Intelligence}, 4\penalty0 (10):\penalty0 828--839, 2022.

\bibitem[Wang et~al.(2021)Wang, Liu, Foster, Chang, Kettimuthu, and Kotamarthi]{wang2021fast}
Jiali Wang, Zhengchun Liu, Ian Foster, Won Chang, Rajkumar Kettimuthu, and V~Rao Kotamarthi.
\newblock Fast and accurate learned multiresolution dynamical downscaling for precipitation.
\newblock \emph{Geoscientific Model Development}, 14\penalty0 (10):\penalty0 6355--6372, 2021.

\bibitem[Sha et~al.(2020)Sha, Gagne~II, West, and Stull]{sha2020deep}
Yingkai Sha, David~John Gagne~II, Gregory West, and Roland Stull.
\newblock Deep-learning-based gridded downscaling of surface meteorological variables in complex terrain. part ii: Daily precipitation.
\newblock \emph{Journal of Applied Meteorology and Climatology}, 59\penalty0 (12):\penalty0 2075--2092, 2020.

\bibitem[Sharma and Mitra(2022)]{sharma2022resdeepd}
Sumanta Chandra~Mishra Sharma and Adway Mitra.
\newblock Resdeepd: A residual super-resolution network for deep downscaling of daily precipitation over india.
\newblock \emph{Environmental Data Science}, 1:\penalty0 e19, 2022.

\bibitem[Groenke et~al.(2020)Groenke, Madaus, and Monteleoni]{groenke2020climalign}
Brian Groenke, Luke Madaus, and Claire Monteleoni.
\newblock Climalign: Unsupervised statistical downscaling of climate variables via normalizing flows.
\newblock In \emph{Proceedings of the 10th International Conference on Climate Informatics}, pages 60--66, 2020.

\bibitem[Winkler et~al.(2024)Winkler, Harder, and Rolnick]{winkler2024climate}
Christina Winkler, Paula Harder, and David Rolnick.
\newblock Climate variable downscaling with conditional normalizing flows.
\newblock \emph{arXiv preprint arXiv:2405.20719}, 2024.

\bibitem[Xiang et~al.(2022)Xiang, Xiang, Guan, Zhang, Zhao, and Zhang]{xiang2022novel}
Li~Xiang, Jie Xiang, Jiping Guan, Fuhan Zhang, Yanling Zhao, and Lifeng Zhang.
\newblock A novel reference-based and gradient-guided deep learning model for daily precipitation downscaling.
\newblock \emph{Atmosphere}, 13\penalty0 (4):\penalty0 511, 2022.

\bibitem[Ling et~al.(2024)Ling, Lu, Luo, Bai, Behera, Jin, Pan, Jiang, and Yamagata]{ling2024diffusion}
Fenghua Ling, Zeyu Lu, Jing-Jia Luo, Lei Bai, Swadhin~K Behera, Dachao Jin, Baoxiang Pan, Huidong Jiang, and Toshio Yamagata.
\newblock Diffusion model-based probabilistic downscaling for 180-year east asian climate reconstruction.
\newblock \emph{npj Climate and Atmospheric Science}, 7\penalty0 (1):\penalty0 131, 2024.

\bibitem[Islam et~al.(2022)Islam, Islam, Shahid, Alam, Biswas, Rahman, Roy, and Kamruzzaman]{islam2022future}
HM~Touhidul Islam, ARMT Islam, Shamsuddin Shahid, GM~Monirul Alam, Jatish~Chandra Biswas, Md~Mizanur Rahman, Dilip~Kumar Roy, and Mohammad Kamruzzaman.
\newblock Future precipitation projection in bangladesh using simclim climate model: a multi-model ensemble approach.
\newblock \emph{Int. J. Climatol}, 42:\penalty0 6716--6740, 2022.

\bibitem[Wu et~al.(2022)Wu, Zhang, Crabbe, and Chandra~Das]{wu2022statistical}
Yichen Wu, Zhihua Zhang, M~James~C Crabbe, and Lipon Chandra~Das.
\newblock Statistical learning-based spatial downscaling models for precipitation distribution.
\newblock \emph{Advances in Meteorology}, 2022\penalty0 (1):\penalty0 3140872, 2022.

\bibitem[Pour et~al.(2018)Pour, Shahid, Chung, and Wang]{pour2018model}
Sahar~Hadi Pour, Shamsuddin Shahid, Eun-Sung Chung, and Xiao-Jun Wang.
\newblock Model output statistics downscaling using support vector machine for the projection of spatial and temporal changes in rainfall of bangladesh.
\newblock \emph{Atmospheric research}, 213:\penalty0 149--162, 2018.

\bibitem[Funk et~al.(2015)Funk, Peterson, Landsfeld, Pedreros, Verdin, Shukla, Husak, Rowland, Harrison, Hoell, et~al.]{funk2015climate}
Chris Funk, Pete Peterson, Martin Landsfeld, Diego Pedreros, James Verdin, Shraddhanand Shukla, Gregory Husak, James Rowland, Laura Harrison, Andrew Hoell, et~al.
\newblock The climate hazards infrared precipitation with stations—a new environmental record for monitoring extremes.
\newblock \emph{Scientific data}, 2\penalty0 (1):\penalty0 1--21, 2015.

\bibitem[Hersbach et~al.(2020)Hersbach, Bell, Berrisford, Hirahara, Hor{\'a}nyi, Mu{\~n}oz-Sabater, Nicolas, Peubey, Radu, Schepers, et~al.]{hersbach2020era5}
Hans Hersbach, Bill Bell, Paul Berrisford, Shoji Hirahara, Andr{\'a}s Hor{\'a}nyi, Joaqu{\'\i}n Mu{\~n}oz-Sabater, Julien Nicolas, Carole Peubey, Raluca Radu, Dinand Schepers, et~al.
\newblock The era5 global reanalysis.
\newblock \emph{Quarterly Journal of the Royal Meteorological Society}, 146\penalty0 (730):\penalty0 1999--2049, 2020.

\bibitem[Mu{\~n}oz-Sabater et~al.(2021)Mu{\~n}oz-Sabater, Dutra, Agustí-Panareda, Albergel, Arduini, Balsamo, Boussetta, Choulga, Harrigan, Hersbach, Martens, Miralles, Piles, Rodríguez-Fernández, Zsoter, Buontempo, and Thépaut]{munoz2021era5}
J.~Mu{\~n}oz-Sabater, E.~Dutra, A.~Agustí-Panareda, C.~Albergel, G.~Arduini, G.~Balsamo, S.~Boussetta, M.~Choulga, S.~Harrigan, H.~Hersbach, B.~Martens, D.~G. Miralles, M.~Piles, N.~J. Rodríguez-Fernández, E.~Zsoter, C.~Buontempo, and J.-N. Thépaut.
\newblock Era5-land: a state-of-the-art global reanalysis dataset for land applications.
\newblock \emph{Earth System Science Data}, 13\penalty0 (9):\penalty0 4349–4383, 2021.
\newblock \doi{10.5194/essd-13-4349-2021}.

\bibitem[Haarsma et~al.(2016)Haarsma, Roberts, Vidale, Senior, Bellucci, Bao, Chang, Corti, Fu{\v{c}}kar, Guemas, et~al.]{haarsma2016high}
Reindert~J Haarsma, Malcolm~J Roberts, Pier~Luigi Vidale, Catherine~A Senior, Alessio Bellucci, Qing Bao, Ping Chang, Susanna Corti, Neven~S Fu{\v{c}}kar, Virginie Guemas, et~al.
\newblock High resolution model intercomparison project (highresmip v1. 0) for cmip6.
\newblock \emph{Geoscientific Model Development}, 9\penalty0 (11):\penalty0 4185--4208, 2016.

\bibitem[Ravela and McLaughlin(2007)]{ravela2007fast}
Sai Ravela and Dennis McLaughlin.
\newblock Fast ensemble smoothing.
\newblock \emph{Ocean Dynamics}, 57\penalty0 (2):\penalty0 123--134, 2007.

\bibitem[Ravela et~al.(2010)Ravela, Marshall, Hill, Wong, and Stransky]{ravela2010realtime}
Sai Ravela, John Marshall, Chris Hill, Andrew Wong, and Scott Stransky.
\newblock A realtime observatory for laboratory simulation of planetary flows.
\newblock \emph{Experiments in fluids}, 48\penalty0 (5):\penalty0 915--925, 2010.

\bibitem[Trautner et~al.(2020)Trautner, Margolis, and Ravela]{trautner2020informative}
Margaret Trautner, Gabriel Margolis, and Sai Ravela.
\newblock Informative neural ensemble kalman learning.
\newblock \emph{arXiv:2008.09915}, 2020.

\bibitem[Goodfellow et~al.(2014)Goodfellow, Pouget-Abadie, Mirza, Xu, Warde-Farley, Ozair, Courville, and Bengio]{goodfellow2014generative}
Ian Goodfellow, Jean Pouget-Abadie, Mehdi Mirza, Bing Xu, David Warde-Farley, Sherjil Ozair, Aaron Courville, and Yoshua Bengio.
\newblock {Generative Adversarial Nets}.
\newblock In \emph{Advances in Neural Information Processing Systems}, volume~27, 2014.

\bibitem[Jolicoeur-Martineau(2018)]{jolicoeur2018relativistic}
Alexia Jolicoeur-Martineau.
\newblock The relativistic discriminator: a key element missing from standard gan.
\newblock \emph{arXiv:1807.00734}, 2018.

\bibitem[Fritsche et~al.(2019)Fritsche, Gu, and Timofte]{fritsche2019frequency}
Manuel Fritsche, Shuhang Gu, and Radu Timofte.
\newblock Frequency separation for real-world super-resolution.
\newblock In \emph{2019 IEEE/CVF International Conference on Computer Vision Workshop (ICCVW)}, pages 3599--3608. IEEE, 2019.

\bibitem[Rodgers(1976)]{rodgers1976retrieval}
Clive~D Rodgers.
\newblock Retrieval of atmospheric temperature and composition from remote measurements of thermal radiation.
\newblock \emph{Reviews of Geophysics}, 14\penalty0 (4):\penalty0 609--624, 1976.

\bibitem[Saha and Ghosh(2019)]{saha2019can}
Anamitra Saha and Subimal Ghosh.
\newblock Can the weakening of indian monsoon be attributed to anthropogenic aerosols?
\newblock \emph{Environmental Research Communications}, 1\penalty0 (6):\penalty0 061006, 2019.

\bibitem[Ravela(2003)]{ravela2003multi}
Srinivas~S Ravela.
\newblock \emph{On multi-scale differential features and their representations for image retrieval and recognition}.
\newblock PhD thesis, University of Massachusetts Amherst, 2003.

\bibitem[Ravela(2014)]{ravela2014spatial}
Sai Ravela.
\newblock Spatial inference for coherent geophysical fluids by appearance and geometry.
\newblock In \emph{IEEE Winter Conference on Applications of Computer Vision}, pages 925--932. IEEE, 2014.

\bibitem[Ravela(2015)]{ravela2015dynamic}
Sai Ravela.
\newblock Dynamic data-driven deformable reduced models for coherent fluids.
\newblock \emph{Procedia Computer Science}, 51:\penalty0 2464--2473, 2015.

\end{thebibliography}
\end{document}